\documentclass{elsart1p}
\usepackage{graphicx}
\usepackage{amssymb}
\begin{document}
\begin{frontmatter}
%
%
%
\title{Charm and Hidden Charm Scalar Resonances \\
in Nuclear Matter}
%
%
\author[label1,label2]{Laura Tol\'os}\ead{tolos@kvi.nl}
\author[label3]{,Raquel Molina}
\author[label3]{,Daniel Gamermann} 
\author[label3]{and Eulogio Oset}
\address[label1]{FIAS, Goethe-Universit\"at Frankfurt am Main, \\ Ruth-Moufang-Str. 1, 60438 Frankfurt am Main, Germany}
\address[label2]{Theory Group, KVI, University of Groningen,  \\ Zernikelaan 25, 9747 AA Groningen, The Netherlands}

\address[label3]{Departamento de F\'{\i}sica Te\'orica and IFIC,
Centro Mixto Universidad de Valencia-CSIC, \\
Institutos de Investigaci\'on de Paterna, Aptdo. 22085, 46071 Valencia, Spain}
\begin{abstract}
We study the properties of the scalar charm resonances
$D_{s0}(2317)$ and  $D_0(2400)$, and the theoretical hidden charm state  $X(3700)$ in nuclear matter.
We find that for the $D_{s0}(2317)$ and $X(3700)$ resonances, with negligible and small width at zero density, respectively, the
width becomes about $100 ~{\rm MeV}$ and $200~{\rm MeV}$ at normal nuclear matter density, accordingly. For $D_0(2400)$ the change in width is relatively less important.
We discuss the origin of this new width and trace it to reactions occurring in 
the nucleus. We also propose a possible experimental test for those modifications in nuclear matter, which  will bring valuable information on the nature of those scalar
resonances and the interaction of $D$ mesons with nucleons. 
\end{abstract}
\end{frontmatter}
%

The study of the properties of elementary particles in nuclei helps to learn about not only the excitation mechanisms in the nucleus but also the properties of those particles. In particular, the renormalization of the properties of scalar mesons in nuclear matter, such as  $\sigma(600)$, $f_0(980)$ or $a_0(980)$, can give some information about their nature, whether they are $q\bar{q}$ states, molecules, mixtures of $q\bar{q}$ with meson-meson components, or dynamically generated resonances resulting from the interaction of two pseudoscalars.

On the other hand, the charm degree of freedom has become a recent topic of analysis. The future FAIR (Facility of Antiproton and Ion Research) project at GSI \cite{FAIR} will investigate, among others, the  modification of the properties of open and hidden charm mesons in a dense baryonic environment.

\begin{figure}[htb]
\begin{center}
\includegraphics[height=5.2 cm, width=4.4 cm]{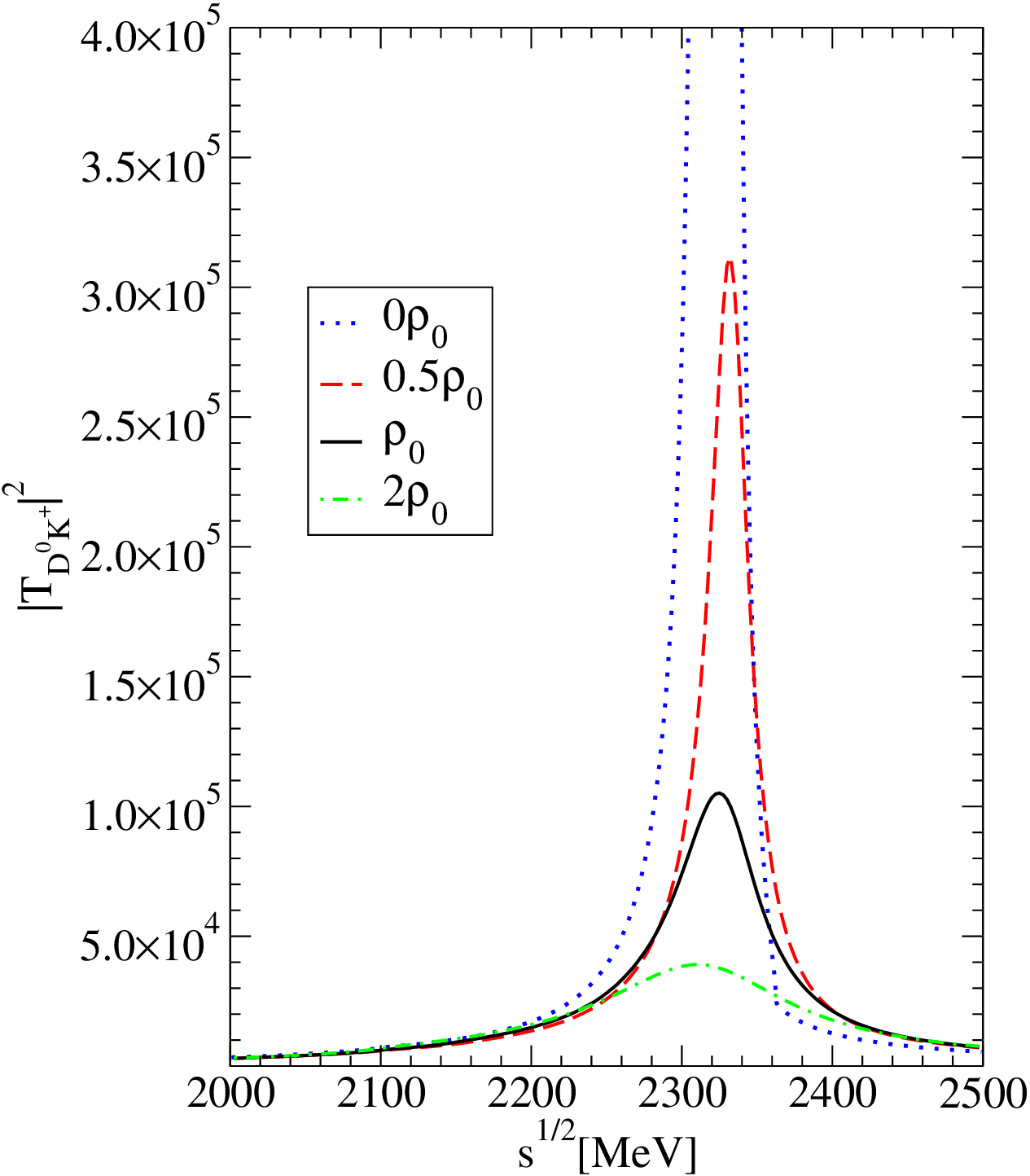}
\hfill
\includegraphics[height=5.2 cm, width=4.4 cm]{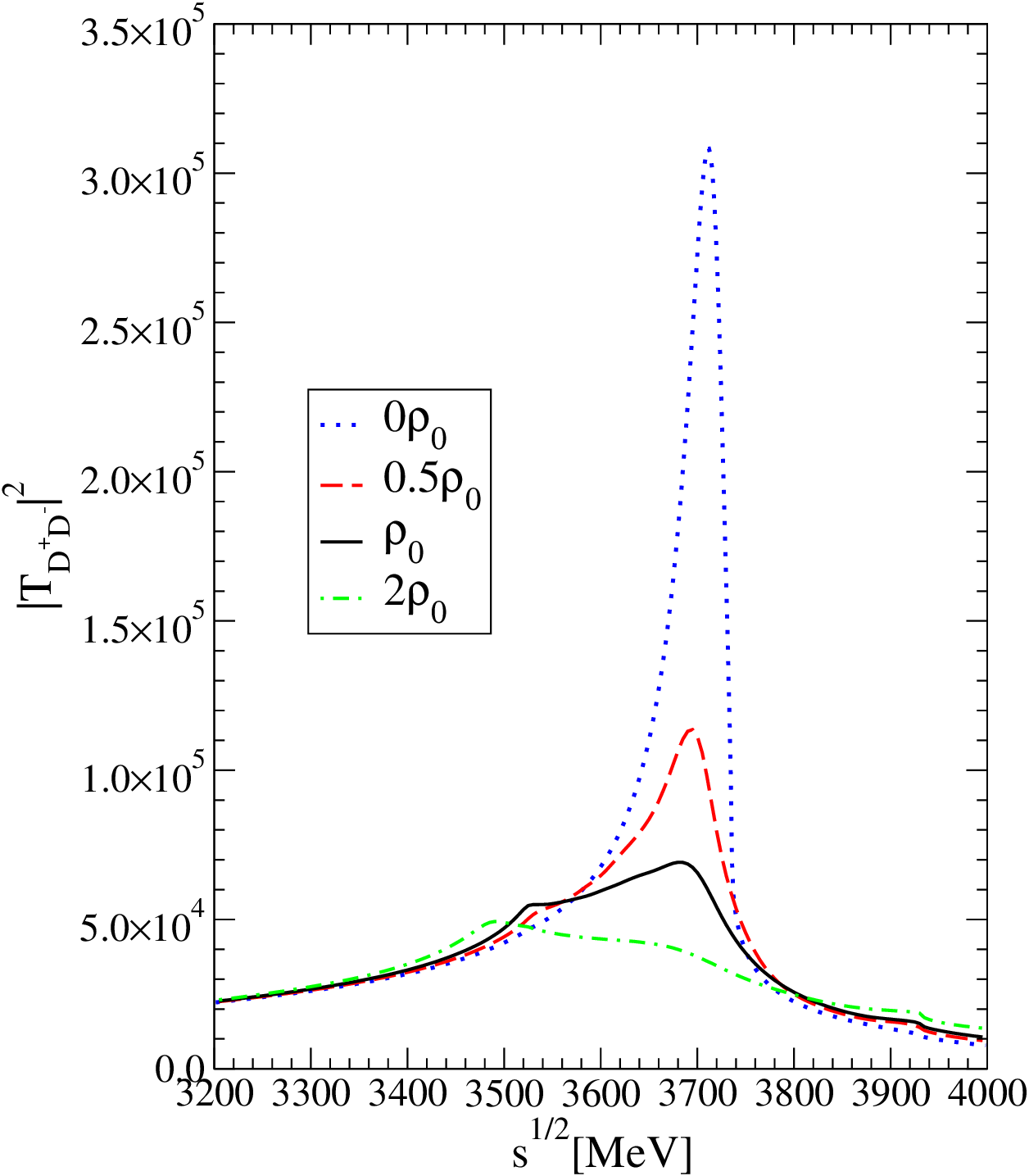}
\hfill
\includegraphics[height=5.2 cm, width=4.4 cm]{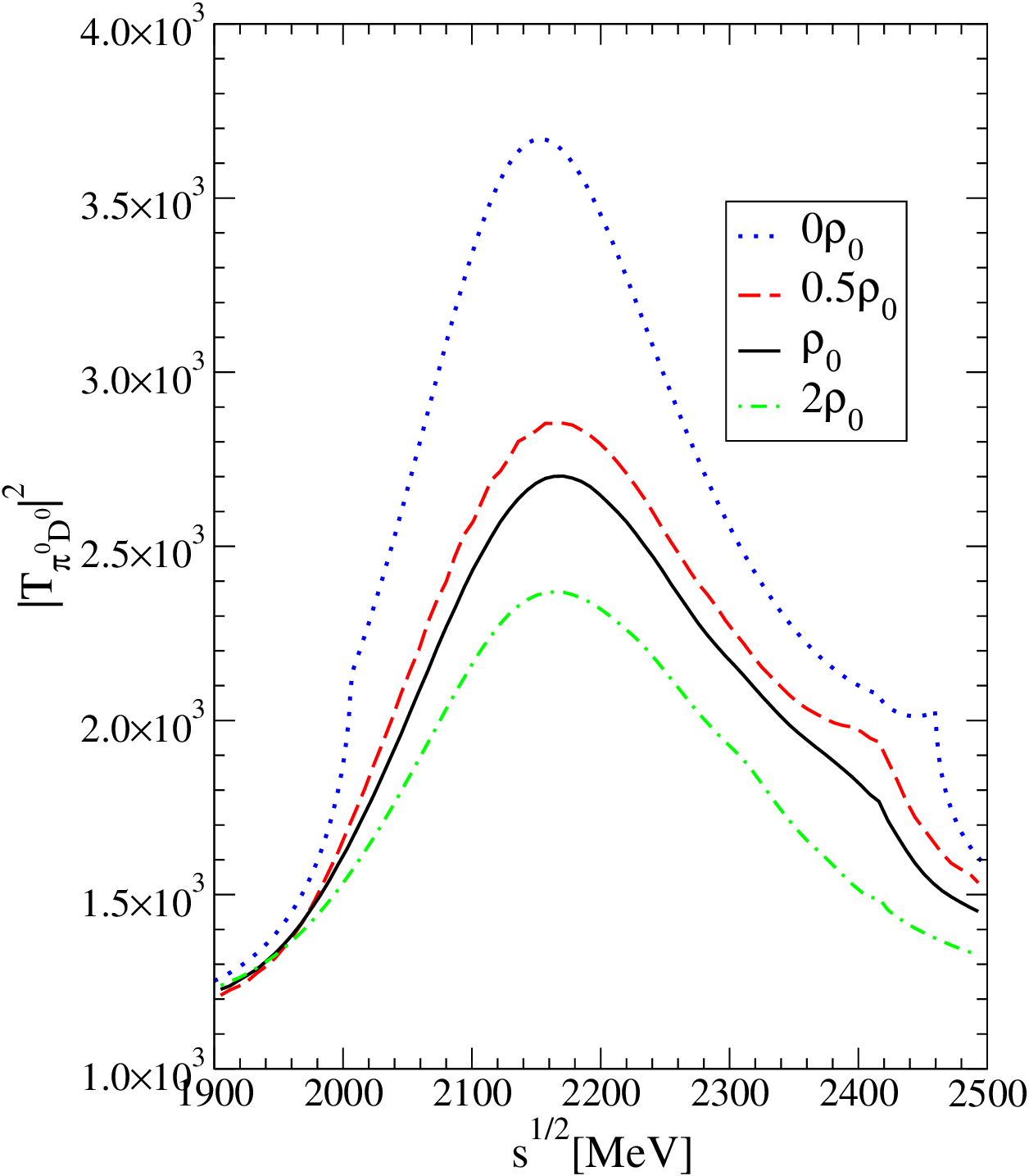}
\caption{$D_{s0}(2317)$ (left), $X(3700)$ (middle) and $D_0(2400)$ (right) resonances in nuclear matter.} \label{fig2}
\end{center}
\end{figure}

In this paper we study the medium modifications of scalar resonances, which are dynamically generated, in the charm sector. Concretely, we study the renormalized properties in nuclear matter of the charm resonances $D_{s0}(2317)$ and $D(2400)$ \cite{Kolomeitsev:2003ac,Hofmann:2003je,Guo:2006fu,Gamermann:2006nm} together with a hidden charm scalar meson, $X(3700)$, predicted in \cite{Gamermann:2006nm}, which might have been observed by the Belle collaboration \cite{Abe:2007sy} via the reanalysis of \cite{Gamermann:2007mu}.

The charm scalar resonances  $D_{s0}(2317)$ and $D(2400)$ and the predicted hidden charm scalar resonance $X(3700)$ are generated dynamically by solving the coupled-channel Bethe-Salpeter equation for two pseudoscalars \cite{Molina:2008nh}. The kernel is derived from an  extrapolation to $SU(4)$ of the $SU(3)$ chiral Lagrangian used to generate the scalar mesons $\sigma(600)$, $f_0(980)$, $a_0(980)$ and 
 $\kappa(900)$ in the light sector. The $SU(4)$ symmetry is, however, strongly 
 broken, mostly due to the explicit consideration of the masses of the vector 
 mesons exchanged between pseudoscalars \cite{Gamermann:2006nm}. 

The analysis of the transition amplitude close to each resonance, or pole in the complex plane, for the different coupled channels give us information about the coupling of the resonance to a particular channel. The $D_{s0}(2317)$ mainly couples to $DK$ system, while the $D_0(2400)$ to $D\pi$ and, secondly, to $D_s \bar K$.  On the other hand, the  hidden charm state $X(3700)$ couples most strongly to 
$D\bar{D}$. Therefore, any change in the $D$ meson properties in nuclear matter will have an important effect on those  resonances. Those modifications are given by the $D$ meson self-energy in nuclear matter. 

In order to obtain the self-energy of the $D$ meson in nuclear matter, the multichannel Bethe-Salpeter equation is solved taking, as bare interaction, a type of broken SU(4) $s$-wave Tomozawa-Weinberg  interaction  using a cutoff regularization scheme. This cutoff is fixed by reproducing the position and the width of the $I=0$ $\Lambda_c(2593)$ resonance, as done similarly in previous works \cite{TOL04}. As a consequence, a new resonance in the $I=1$ channel, $\Sigma_c(2880)$, is generated \cite{LUT06}. The in-medium solution 
incorporates  Pauli blocking effects, baryon mean-field bindings, and $\pi$ and $D$ meson self-energies in a self-consistent manner (see Ref.~\cite{TOL07}). The $p$-wave self-energy is also included via the corresponding $Y_cN^{-1}$ excitations \cite{Molina:2008nh}, where is taken into account a heavy-form factor in the vertex \cite{Navarra}. 

 In Fig.~\ref{fig2}, the charm scalar resonances $D_{s0}(2317)$ and $D_0(2400)$ as well as the predicted hidden charm one $X(3700)$ are displayed by showing the squared transition amplitude for the corresponding dominant channel at different densities. We obtain that in the case of the $D_{s0}(2317)$ and 
$X(3700)$ resonances, which have a zero and small width, respectively,
the medium effects lead to widths of the order of 100 and 200
MeV at normal nuclear matter density, correspondingly. The origin of this increased width can be traced back to the opening of new many-body decay channels, as the $D$ meson gets absorbed in the nuclear medium via $DN$ and $DNN$ inelastic reactions.  For the $D_0(2400)$, we observe an 
 extra widening from the already large width of the resonance in free space. However,
  the large original width makes the medium effects comparatively much 
  weaker than for the other two resonances \cite{Molina:2008nh}. As for a shift in mass, the changes in mass are small and the uncertainties of our model, such as the way that the $SU(4)$ symmetry is broken and the use of light or heavy meson form factors, do not allow us to extract any clear conclusion.  

The study of the width of those resonances and the medium
reactions contributing to it provides information on the features of the
resonances and the self-energy of the $D$ meson in a nuclear medium. We suggest to look at transparency ratios to investigate those in-medium widths. This magnitude, which gives the survival probability in production reactions in  nuclei, is very sensitive to the absorption rate of any resonance inside nuclei, i.e., to its in-medium width.

 In summary, we have evaluated the renormalized properties in nuclear matter of the charm scalar $D_{s0}(2317)$ and $D(2400)$, and the predicted hidden charm $X(3700)$ resonances. While the $D_{s0}(2317)$ and $X(3700)$ develop a spectacular width of 100 and 200 MeV, respectively, the medium effects on $D(2400)$ are comparatively less important. We conclude that the experimental analysis of those properties is a valuable test of the dynamics of the $D$ meson interaction with nucleons and
nuclei, and the nature of those charm and hidden charm scalar resonances. These are topics which are subject of much debate at present. The results obtained 
here should stimulate experimental work in hadron facilities, in particular at 
FAIR \cite{FAIR}, where the charm degree of freedom will play a leading role.

\section*{Acknowledgments}

L.T. wishes to acknowledge support from Deutsche Forschungsgemeinschaft to attend PANIC08 conference.

%
%
%

%
\end{document}